\documentstyle[12pt]{article}

\def\reference{\parskip 0pt\par\noindent\hangindent 0.5 truecm}

\begin{document}

\title{The 2dF gravitational lens survey}

\author{
	Daniel J.\ Mortlock$^{1,2}$ \and
 	Darren S.\ Madgwick$^{1}$ \and
 	Ofer Lahav$^{1}$
}

\date{}
\maketitle

{\center
	$^1$ Institute of Astronomy, Madingley Road, Cambridge CB3 0HA,
	United Kingdom \\
	mortlock@ast.cam.ac.uk, dsm@ast.cam.ac.uk, lahav@ast.cam.ac.uk\\[3mm]
	$^2$ Astrophysics Group, Cavendish Laboratory, Madingley Road,
	Cambridge CB3 0HE,
        United Kingdom \\
}

\begin{abstract}
The 2 degree Field (2dF) galaxy redshift survey will involve obtaining 
spectra of approximately $2.5 \times 10^5$ objects which have
previously been identified as galaxy candidates on morphological
grounds. 
Included in these spectra should be about ten gravitationally-lensed
quasars, all with low-redshift galaxies as deflectors 
(as the more common lenses
with high-redshift deflectors will be rejected from the survey as multiple
point-sources). 
The lenses will appear as superpositions of 
galaxy and quasar spectra, and either cross-correlation techniques
or principal components analysis should be able to identify 
candidates systematically.
With the 2dF survey approximately half-completed it is now
viable to begin a methodical search for these spectroscopic lenses,
and the first steps of this project are described here.
\end{abstract}


\bigskip

\section{Introduction}
\label{section:intro}

It is possible to discover gravitational lenses spectroscopically
by searching for quasar emission lines in the spectra 
of local galaxies.
This is the only way to find lenses with low-redshift deflectors;
these are particularly valuable as the lens galaxies can be studied
in great detail and the timescales for microlensing are short 
enough to permit a range of interesting measurements 
(Mortlock \& Webster 2000 and references therein).
Kochanek (1992) and Mortlock \& Webster (2000) calculated the
expected event rate, and several tens of spectroscopic lenses
should be forthcoming from the current generation of galaxy redshift
surveys (GRSs).\footnote{The use of existing spectra is critical 
to the efficiency of this method; this is discussed elsewhere in this
issue by Mortlock \& Drinkwater (2001).}

The first large GRS, the Center for Astrophysics (CfA) survey 
(Geller \& Huchra 1989),
resulted in the discovery of the lens Q~2237+0305 (Huchra et al.\ 1985), 
although it was
an essentially serendipitous event.
There have been no other quasar lenses discovered in this manner,
but several searches for lensed emission line galaxies have been
undertaken.
The only confirmed spectroscopic discoveries were the 
two Einstein rings found 
by Warren et al.\ (1996) and Hewett et al.\ (1999) in their sample
of large elliptical galaxies.
Even though there are only several hundred galaxies in the survey,
the regularity of their spectra is such that a very sensitive search
for discrepant emission lines is possible (Willis et al.\ 2001);
several more candidates have been identified and are awaiting
re-observation. 
Hall et al.\ (2000) also announced the identification of 
several candidate lenses in the Canadian Network for Observational Cosmology 
(CNOC) GRS (Yee et al.\ 2000), although these too are yet to be 
confirmed.
The next generation of galaxy survey is exemplified by
the Sloan Digital Sky Survey (SDSS; York et al.\ 2000) which, with $\sim 10^6$ 
galaxy spectra, should yield at least 50 spectroscopic lenses
(not to mention the several hundred lenses expected to be resolved
in the imaging survey). The SDSS
has only recently got under way, but, at the time of writing, 
over $5 \times 10^4$ spectra have been obtained, and the 
pipeline reduction software should be able to select
lens candidates with some reliability, although none have been
announced so far.

This paper details the beginnings of a lens survey based on the spectra
taken as part of the largest existing galaxy sample,
the 2 degree Field (2dF) GRS (Section~\ref{section:grs}).
After a brief discussion of the lens statistics 
(Section~\ref{section:stat}), 
the possible spectral analysis methods are examined 
(Section~\ref{section:gls}).
These points are summarised in Section~\ref{section:conc}.

\section{The 2dF galaxy redshift survey}
\label{section:grs}

The 2dF GRS
(e.g., Folkes et al.\ 1999; Cole et al.\ 2001) is, formally,
a spectroscopic survey of $\sim 2.5 \times 10^5$
extended images, all of which have isophotal magnitudes brighter
than $B_{\rm J} = 19.45$. 
The data are obtained using the 2dF instrument, an 
automatically-configurable multi-fibre spectrograph on the 
Anglo-Australian Telescope (AAT).
The $\sim 10^5$ spectra obtained so far have a resolution of about
10 \AA, and cover the range from $\sim 3700$ \AA\ to $\sim 8100$ \AA.
Whilst the wavelength calibration is 
excellent, there have been some problems with the flux calibration,
and so continuum levels remain quite uncertain. 
The spectra are typically the result of 45-min integrations, 
and so the signal-to-noise ratio in each pixel is 
between $\sim 10$ and $\sim 20$ for an object at the survey limit.

It is expected that $\sim 95$ per cent of the objects selected for 
the survey will be local galaxies, with the remainder being mainly 
misidentified Galactic stars. 
However a few of the survey galaxies will be multiply-imaging
background quasars, and, if the quasars are sufficiently bright, 
should be identifiable as gravitational lenses.
Two main questions present themselves at this point: how many lenses
might be detectable in these spectra, and what is the best way to 
find any lenses that might be present? 
The answers are related, in so far as any given search method is linked 
to the effective ``depth'' of the lens survey, but it is possible 
to obtain an estimate of the number of lenses without fully understanding
how to find them. 

\section{Lens statistics}
\label{section:stat}

Kochanek (1992) and Mortlock \& Webster (2000) applied standard 
techniques to determine the rate of lensed quasars expected in a 
generic GRS, and a more detailed calculation verified these 
results for the 2dF survey
(Mortlock \& Webster 2001).
The number of lenses depends principally
on two factors:
the depth of the survey and the quality of the spectra. 
The latter, potentially a somewhat ambiguous notion, 
can be characterised in terms 
of $\Delta m_{\rm qg}$, which is defined as follows 
(Kochanek 1992):
for a lens in which the galaxy has magnitude $m_{\rm g}$ and 
the quasar images have total magnitude $m_{\rm q}$, 
the presence of the quasar emission lines is detectable if
$m_{\rm q} \leq m_{\rm g} + \Delta m_{\rm qg}$, but the quasar
is undetectable if it is fainter than this.
The value of $\Delta m_{\rm qg}$ clearly increases with the 
integration time of the observations, but is also critically 
dependent on the properties of galaxy and quasar spectra. 
Kochanek (1992) estimated $\Delta m_{\rm qg} \simeq 4$ for 
the SDSS spectra, and Mortlock \& Webster (2000) used 
$\Delta m_{\rm qg} \simeq 2$ for the lower quality 2dF spectra. 
These figures then led to the estimate that the 2dF GRS should yield
about 10 lensed quasars (Mortlock \& Webster 2000), although it
could be as low as $\sim 5$ (if $\Delta m_{\rm qg} \simeq 1$) 
or as high as $\sim 20$ (if $\Delta m_{\rm qg} \simeq 3$).
Fortunately the value of $\Delta m_{\rm qg}$ appropriate to a given
search technique can be evaluated by analysing
simulated composite lens spectra.

\section{The 2dF gravitational lens survey}
\label{section:gls}

The process of finding lensed quasars in the 2dF galaxy spectra can 
be split into three quite distinct phases. The first of these is the
identification of candidate lenses on purely spectroscopic grounds. 
The vast majority of the spectra should be uninteresting, at least
in the present application, but a small fraction should 
stand out as having either unusual continua, or, more likely, what
may be broad emission lines. This set of (probably) several
thousand spectra will include white dwarfs; unlensed quasars misidentified
by the APM software; and, with luck, several lensed quasars. The second
phase of the analysis is to examine each spectrum in detail (possibly
by eye) with a view to removing all those objects which are clearly 
not quasars. This step could also profitably include using any other
data which is already available, such as images from the
Digitized Sky Survey 
or the Schmidt plates from which the 2dF objects were selected. Finally, the
(hopefully) small set of strong lens candidates would have to be 
imaged at high resolution ($< 1$ arcsec) with a view to determining
their nature morphologically. In terms of quantifiable resources this
last step is by far the most expensive, as the automatic spectral analysis
should take less than a second per object.
However
the opposite is true when intellectual effort is considered, as the 
manual inspection of the initial candidate list and (especially) the
imaging observations are conceptually straight-forward. 
The complex
task is determining how best to find a subset of the survey spectra that
contains all the lenses and not too much else. 

One approach to this problem would be to work backwards -- if it 
is assumed that all the detectable lenses will primarily have 
broad quasar emission lines then the task reduces to that of identifying
the spectra that have emission line-like features. Cross-correlation
techniques, combined with continuum removal, could certainly detect
some such objects. Unfortunately many galaxy spectra also have features
that could be confused with the broad lines of a fainter, superimposed
spectrum. Further, this is clearly an inefficient procedure, as much of
the available information remains unused. Finally,
such a method becomes increasingly difficult once any attempt is 
made to include non-Gaussian effects such as variation in the continuum
levels and the presence of sky lines.

The method that has been adopted is based on Monte Carlo
simulations of a large number of lenses. 
Using the formalism described in Mortlock \& Webster (2000)
it is possible to generate deflector-source pairs that reflect the 
variation in lensing likelihood with source redshift and deflector
properties. 
A galaxy-quasar pair generated in this way 
is a composite object with only one degree of freedom: the
relative contribution of the two components to the resultant 
spectrum. This is very sensitive to the properties of the lens, 
but such details need not be treated explicitly as the range
of possible weightings can be analysed without knowing the 
actual image configurations. 
The most important aspect of this method is that,
for a given galaxy-quasar pair, 
a very realistic composite spectrum can be generated by simply combining
real 2dF galaxy and quasar spectra. The noise properties of the 
spectra are certainly correct provided both spectra are kept in their
observed frame; the one possible important omission is reddening of 
the background quasar by dust in the lens.
This too can be included in the simulations, but it would 
incur the penalty of parameterisation; this process is otherwise 
model-independent.
A promising candidate for the analysis method is 
principal components analysis (PCA; Murtagh \& Hecht 1987), 
which is already being used extensively within the 2dF collaboration,
and is well suited to the task at hand. The survey galaxies 
cover a definite region in the multi-dimensional space of component
coefficients (e.g., Folkes et al.\ 1999) and the lenses should 
inhabit a contiguous region, with the distance from the galactic 
locus increasing with the brightness of the quasar component.
Unfortunately this process has not
yet been implemented; the central ideas are quite clear, but 
quantitative estimates of the completeness and efficiency of the lens
survey cannot yet be made.

\section{Conclusions}
\label{section:conc}

It should be possible to find $\sim 10$ gravitationally-lensed quasars 
in the 2dF GRS spectra.
Cross-correlation techniques can be used to search
for quasars' emission lines, but a more powerful (and more simply implemented)
scheme is to generate large numbers of simulated lens spectra 
and to then analyse them using existing software. 
If the quasar and galaxy spectra used are taken from the 2dF samples
this method has the enormous advantage of automatically including 
various subtleties of the survey data, without the need for detailed
modelling. By identifying the regions of parameter space inhabited
by these spectra, it is possible to characterise the sensitivity of
the lens search (effectively measuring the $\Delta m_{\rm qg}$ 
parameter discussed in Section~\ref{section:stat}). More importantly,
the real spectra that fall into the same regions of
parameter space are immediately 
identifiable as candidate lenses. Implementation of these techniques
should bring the first positive results within the next year.

\section*{Acknowledgements}

Thanks to Rachel Webster for first suggesting this avenue of 
investigation, and also to 
the entire 2dF survey team; their support will be critical to the success of
this project.
DJM is funded by PPARC.

\section*{References}

\reference Cole S., et al., 2001, MNRAS, submitted
	(astro-ph/0012429)
\reference Folkes, S.R., et al., 1999, MNRAS, 308, 459
\reference Geller, M.J., Huchra, J.P., 1989, Science, 246, 897
\reference Hall, P.B., et al., 2000, AJ, 120, 1660
\reference Hewett, P.C., Warren, S.J., Willis, J.P., Bland-Hawthorn, J.,
	Lewis, G.F., 1999, in 
	Imaging the Universe in Three Dimensions,	
	eds.\ van Breugel, W., Bland-Hawthorn, J., ASP, 94
\reference Huchra, J.P., Gorenstien, M., Kent, S., Shapiro, I., Smith, G.,
        Horine, E., Perley, R., 1985, AJ, 90, 691
\reference Kochanek, C.S., 1992, ApJ, 397, 381
\reference Mortlock, D.J., Drinkwater, M.J., 2001, PASA, in press
\reference Mortlock, D.J., Webster, R.L., 2000, MNRAS, 319, 879
\reference Mortlock, D.J., Webster, R.L., 2001, MNRAS, 321, 629
\reference Murtagh, F., Hecht, A., 1987, Multivariate Data Analysis,
	Reidel
\reference Warren, S.J., Hewett, P.C., Lewis, G.F., M{\o}ller, P., Iovino, A.,
        Shaver, P.A., 1996b, MNRAS, 278, 139
\reference Willis, J.P., Hewett, P.C., Warren, S.J., Lewis, G.F., 
	2001, MNRAS, submitted
\reference Yee, H.K.C., et al., 2000, ApJS, 129, 475
\reference York, D.G., et al., 2000, AJ, 120, 1579

\end{document}